\documentclass[prl,showpacs,floatfix,epsfig,twocolumn,superscriptaddress]{revtex4}

\usepackage{graphicx}
\usepackage{dcolumn}
\usepackage{bm}

\begin{document}
\title{Dimensional crossover and anomalous magnetoresistivity in single crystals $Na_xCoO_2$}
\author{C. H. Wang} \affiliation{Hefei National Laboratory
for Physical Science at Microscale and Department of Physics,
University of Science and Technology of China, Hefei, Anhui
230026, People's Republic of China}

\author{X. H. Chen} \altaffiliation{Corresponding author: chenxh@ustc.edu.cn} \affiliation{Hefei National Laboratory for
Physical Science at Microscale and Department of Physics,
University of Science and Technology of China, Hefei, Anhui
230026, People's Republic of China}

\author{J. L. Luo} \affiliation{Beijing National Laboratory for Condensed Matter
Physics, Institute of Physics, Chinese Academy of Science, Beijing
100080, People's Republic of China}

\author{G. T. Liu} \affiliation{Beijing National Laboratory for Condensed Matter
Physics, Institute of Physics, Chinese Academy of Science, Beijing
100080, People's Republic of China}

\author {X. X. Lu} \affiliation{Hefei National Laboratory
for Physical Science at Microscale and Department of Physics,
University of Science and Technology of China, Hefei, Anhui
230026, People's Republic of China}
\author {H. T. Zhang} \affiliation{Hefei National Laboratory
for Physical Science at Microscale and Department of Physics,
University of Science and Technology of China, Hefei, Anhui
230026, People's Republic of China}

\author {G. Y. Wang} \affiliation{Hefei National Laboratory
for Physical Science at Microscale and Department of Physics,
University of Science and Technology of China, Hefei, Anhui
230026, People's Republic of China}
\author {X. G. Luo} \affiliation{Hefei National Laboratory
for Physical Science at Microscale and Department of Physics,
University of Science and Technology of China, Hefei, Anhui
230026, People's Republic of China}
\author {N. L. Wang} \affiliation{Beijing National Laboratory for Condensed Matter
Physics, Institute of Physics, Chinese Academy of Science, Beijing
100080, People's Republic of China}

\date{\today}

\begin{abstract}

The in-plane ($\rho_{ab}$) and c-axis ($\rho_c$) resistivities,
and the magnetoresistivity of single crystals $Na_xCoO_2$ with x =
0.7, 0.5 and 0.3 were studied systematically. $\rho_{ab}(T)$ shows
similar temperature dependence between $Na_{0.3}CoO_2$ and
$Na_{0.7}CoO_2$, while $\rho_c(T)$ is quite different. A
dimensional crossover from two to three occurs with decreasing Na
concentration from 0.7 to 0.3. The angular dependence of in-plane
magnetoresistivity for 0.5 sample shows  a \emph{"d-wave-like"}
symmetry at 2K, while the \emph{"p-wave-like"} symmetry at 20 K.
These results give an evidence for existence of a \emph{spin
ordering orientation} below 20 K turned by external field, like
the stripes in cuprates.

\end{abstract}


\maketitle
\newpage

Recently the study of layered sodium cobaltate oxide $Na_xCoO_2$
has became one of the hot topics in research of condensed matter
physics. Na doping leads to the change of spins from spin 1/2 for
$Co^{4+}$ to spinless for $Co^{3+}$. The discovery of
superconductivity with $T_c\sim$ 5 K in
$Na_{0.35}CoO_2\cdot1.3H_2O$\cite{Takada} makes one consider that
$Na_xCoO_2$ may be a system like cuprates where superconductivity
occurs in a doped Mott insulator. Furthermore, it is also expected
that the triangle lattice based cobaltates may exhibit some novel
electronic and magnetic phases, for example, the Anderson's RVB
states\cite{Anderson} and the strong topological frustration
phases\cite{Baskaran,Kumar,Qiang-Hua Wang}. The in-plane magnetic
correlation of $Na_xCoO_2$ is complex and still not clear. The
$^{59}Co$ nuclear quadrupolar resonance measurement suggests a
two-dimensional antiferromagnetic correlation in $Na_{0.35}CoO_2
\cdot yH_2O$\cite{Fujimoto}. However, the band structure
calculations predict a ferromagnetic spin fluctuations within the
$CoO_2$ plane for $Na_xCoO_2$\cite{Singh1,Singh2}. The existence
of in-plane ferromagnetic correlations is confirmed by inelastic
neutron scattering experiments for x=0.75\cite{Boothroyd} and
x=0.82\cite{keimer} crystals.

Besides the magnetic properties, $Na_xCoO_2$ system also shows
many anomalous transport properties. Extremely large and magnetic
field dependent thermoelecreic power (TEP) is observed. The
enhancement of TEP is believed to be due to spin entropy in
$Na_{0.68}CoO_2$ \cite{wangyayu01}. The Hall coefficient is found
to be linear temperature-dependent in a wide temperature range and
shows no saturation up to 500 K\cite{wangyayu02}, while in a
conventional metal Hall coefficient is temperature independent.
The unusual linear-T resistivity is found in low
temperatures\cite{wangyayu01,foo,arpes-nacoo} showing non-Fermi
liquid behavior. While observation of a $T^2$ dependence of
resistivity at ultra low temperature and the satisfaction of the
Wiedemann-Franz law in $Na_{0.7}CoO_2$ show the validity of
ordinary Fermi-liquid state in ultra-low temperatures with
enormous e-e sacttering\cite{lisy}. In addition, a so-called
"incoherent-coherent" transition with decreasing temperature is
observed in $\rho_c(T)$. ARPES experiments reveal that
well-defined quasiparticle peaks develop only at low temperature
in $Na_{0.75}CoO_2$ when a dimensional crossover
occurs\cite{arpes-nacoo,Valla}. The electronic and magnetic
properties in $Na_xCoO_2$ is very sensitive to the Na
content\cite{Motohashi}. Therefore, it would be helpful to study
the evolution of transport properties with doping level in this
system.

The anisotropic transport measurement is a useful way to the
understanding of electronic and magnetic properties such as the
dimensional transition. Magnetoresistance measurement is also a
useful tool to get further insight into the anomalous charge
transport since it is more sensitive to the change in the charge
carrier scattering rate $1/\tau$, effective mass $m^*$ and the
geometry of the Fermi surface. In this letter, the anisotropic
transport properties of $Na_{x}CoO_2$ single-crystals are studied
systematically. A dimensional crossover from two to three occurs
with decreasing the Na content. The angular dependence of the
in-plane resistivity for $Na_{0.5}CoO_2$ shows a two-fold symmetry
at 20 K, while four-fold symmetry at 2 K. Such anomalous angular
dependence of MR is related to a special \emph{"spin ordering
orientation"}, such as the stripes, which can be turned by the
applied field.

High quality single crystals $Na_{0.7}CoO_2$ were grown using the
flux method. The typical dimensional is about $2 \times 1.5 \times
0.01 mm^3$ with the shortest dimension along the c axis. The x=0.5
and 0.3 samples were achieved by Na deintercalation of
$Na_{0.7}CoO_2$ in solutions of iodine- and bromine-dissolved
acetonitrile. The actual Na concentration is determined by
inductively coupled plasma analysis (ICP) experiments. No impure
phase is detected within experimental error of 2\% in x-ray
diffraction measurements. The c-axis lattice parameter is about
10.960, 11.115 and 11.215 $\AA$ for x=0.7, 0.5 and 0.3,
respectively. This is consistent with the data reported by Foo et
al\cite{foo}. To obtain the coincident results, one single crystal
was cut into two pieces. One is for in-plane resistivity
measurement and another for out-of-plane measurement. The
resistivity and magnetoresistance were performed in Quantum Design
PPMS systems.

$\rho_{ab}(T)$ and $\rho_c(T)$ for $Na_{x}CoO_2$ with x=0.7, 0.5,
and 0.3 are presented in Fig. 1(a) and (b). As reported by other
groups\cite{wangyayu01,foo,arpes-nacoo}, a T-linear behavior of
$\rho_{ab}(T)$ is observed in $Na_{0.7}CoO_2$ in low temperatures
below 100 K, indicating non-Feimi-liquid-like
behavior\cite{arpes-nacoo}. In the temperature range between 100
and 180 K, $\rho_{ab}(T)$ can be well fitted by a $T^{3/2}$ law.
Above 180 K, $\rho_{ab}(T)$ follows another power-law ($T^\alpha$,
$\alpha <1$) behavior. It is worth noting that the
$T^{3/2}$-dependent behavior has been reported by Rivadulla et
al.\cite{Rivadulla} in several metallic systems including
$Na_{x}CoO_2$. They argued that this phenomenon presents in metals
which approach the Mott-Hubbard transition from the
itinerant-electron side, and concluded that locally cooperative
bond-length fluctuations at isolated clusters can account for this
behavior.
\begin{figure}[t]
\centering
\includegraphics[width=9cm]{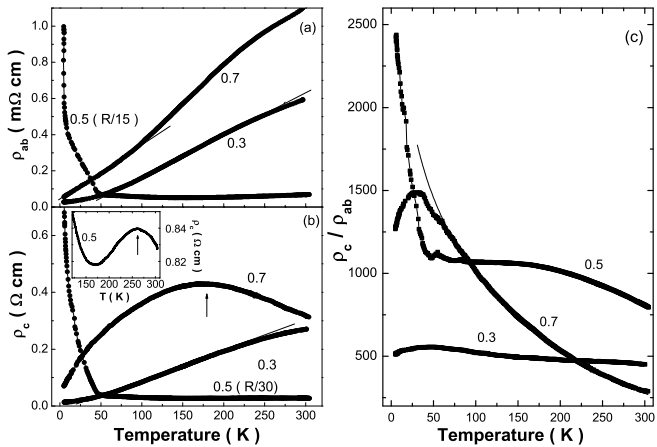} \vspace*{-5mm}
\caption{\label{fig:epsart} Temperature dependence of (a) in-plane
 resistivity ($\rho_{ab}$); (b) out-of-plane
resistivity ($\rho_c$),  and the thin line guides for eyes; (c)
temperature dependence of anisotropy ($\rho_{ab}/\rho_c$) for
$Na_{x}CoO_2$ with x=0.7, 0.5, and 0.3, the line is the fitted
result by $ln(T_0/T)$ for x=0.7 sample.} \vspace*{-5mm}
\end{figure}
$\rho_{ab}(T)$ for x=0.5 sample shows a weak metallic behavior
above 160 K, then a weak insulating behavior as T falls toward 50
K. Below 50 K, $\rho_{ab}(T)$ grows sharply, indicating the
opening of charge ordering gap at about 50K. $\rho_{ab}(T)$
increases more sharply below 20 K. The result is consistent with
that observed by Foo et al\cite {foo}.

We have carefully analyzed the data of 0.3 sample. It is found
that the $T^2$ behavior is observed in $\rho_{ab}(T)$ below about
30 K as previous report\cite{foo}. A $T^{3/2}$ temperature
dependence is observed from 30 to about 100 K, then a T-linear
behavior between about 100 and 220 K. Above 220 K, $\rho_{ab}(T)$
can be described by $T^\alpha$' with $\alpha$'$<$1 as the case of
$Na_{0.7}CoO_2$. In comparison with x=0.7 sample, $\rho_{ab}(T)$
for 0.3 sample shows also four different temperature dependence
regimes. But the $T^2$ dependence of resistivity occurs below 30
K, which is much higher than that (1 K) observed in the 0.7
sample\cite{lisy}. In addition, the T-linear behavior begins to
show up at 220 K, while at 100 K in the 0.7 sample. Although
$\rho_{ab}(T)$ experiences similar temperature dependence with
increasing temperture ($T^2 \rightarrow T^{3/2}\rightarrow T
\rightarrow$$T^\alpha$'($\alpha$'$<$1)) to the case of
$Na_{0.7}CoO_2$ ($T^2\rightarrow T\rightarrow T^{3/2}\rightarrow$
$T^\alpha(\alpha<1)$), $\rho_{c}(T)$ shows quite different
behavior.

The temperature dependences of $\rho_{c}$ for the three samples
are shown in Fig. 1(b). $\rho_{c}(T)$ shows
 a so-called "incoherence-coherence" transition peak at $T_M$
around 180 K for $Na_{0.7}CoO_2$ sample and 260 K for $x = 0.5$,
respectively, similar to the report by Valla et
al\cite{Valla,Terasaki} in $(Bi_{0.5}Pb_{0.5})_2Ba_3Co_2O_y$ ($T_M
\sim$200 K) and $NaCo_2O_4$($T_M \sim$180 K). They considered it
as a crossover in the number of effective dimensions from two to
three. However, such maximum in $\rho_{c}(T)$ is not observed in
the 0.3 sample over the whole temperature range.  As observed in
$\rho_{ab}(T)$ for x=0.3 sample, $\rho_{c}(T)$ also shows a
$T^{2}$ dependent behavior in the temperatures below 30 K, and
T-linear behavior from 100 to 220 K. This indicates that the x=0.3
sample has the same temperature dependence for both $\rho_{ab}(T)$
and $\rho_c(T)$, suggesting the same scattering mechanism. It may
explain why no "incoherence-coherence" transition peak is
observed. While for x=0.7 sample, the temperature dependence of
$\rho_{ab}(T)$ is completely different from that of $\rho_{c}(T)$.
It suggests that there exists different scattering mechanism
between in-plane and out-of plane charge transports. For
$Na_{0.5}CoO_2$, the $\rho_{c}(T)$ shows an
"incoherence-coherence" transition peak at 260 K, and a metallic
behavior is observed from 260 K to 165 K. Below 165 K, it shows a
similar behavior as observed in $\rho_{ab}$, but the insulating
behavior is much sharper than that in ab-plane. It should be
pointed out that the gross feature in $\rho_{ab}$ and $\rho_c$ is
similar to each other. In one word, $\rho_{ab}(T)$ and $\rho_c(T)$
show the same scattering mechanism for $Na_{0.3}CoO_2$, while
different scattering mechanism for $Na_{0.7}CoO_2$. The
temperature corresponding to "incoherent-coherent" transition
increases with decreasing Na content. These results suggest a
dimensional crossover from two to three with decreasing Na
concentration from 0.7 to 0.3, being similar to the observation in
$La_{2-x}Sr_xCuO_4$ with increasing x\cite{nakamura}. In both of
the cases, the carrier increase leads to a dimensional crossover
from two to three companied with the decrease in spin correlation.

$\rho_c(T)/\rho_{ab}(T)$ is plotted in Fig. 1(c).
$\rho_c(T)/\rho_{ab}(T)$ shows weak temperature dependence for the
x=0.3 sample. However, $\rho_c(T)/\rho_{ab}(T)$ increases with
lowering temperature and has strong temperature dependence above
50 K for the x=0.7 sample, and it can be well fitted by
$ln(T_o/T)$ above 50 K. $\rho_c(T)/\rho_{ab}(T)$ shows also a weak
temperature dependence above 165 K for the x=0.5 sample, and a
saturation between 165 and 50 K, while below 50 K the anisotropy
increases quickly, implying that the insulating behavior of
$\rho_{c}$ is stronger than $\rho_{ab}$ due to charge ordering.
These results further indicate that a dimensional crossover from
two to three occurs with decreasing Na content. It should be
addressed that although $\rho_c$ for the $0.5$ and $0.7$ samples
show a "incoherent-coherent" transition peak at 260 and 180 K,
respectively, the $\rho_c/\rho_{ab}$ does not show any anomaly at
the peak temperature. The anisotropy shows a maximum at about 40 K
for x=0.7 sample, which is much lower than the so-called
coherence-incoherence transition (180 K). For the 0.5 sample, the
anisotropy increases slightly with decreasing temperature, and
shows a saturation at around 165 K which coincides with the
crossover from metallic to nonmetallic behaviors in $\rho_{ab}$
(165 K). It should be pointed out that the low temperature
anisotropy maximum for the x=0.7 and x=0.3 samples corresponds to
the temperatures at which the in-plane thermal conductivity
$\kappa$ and Hall coefficient $R_H$ begin to slightly drop for
both $x = 0.71$ and $0.31$\cite{foo}. It suggests that the
anisotropy maximum is correlated with the change in the charge
transport character.

\begin{figure}[t]
\centering
\includegraphics[width=9cm]{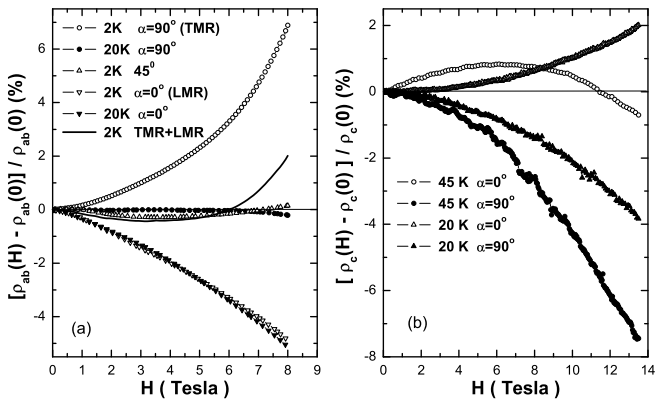}
\vspace*{-6mm} \caption{\label{fig:epsart} In-plane and
out-of-plane magnetoresistivities with different angles between H
and I for $Na_{0.5}CoO_2$ at 2, 20, and 45 K, respectively. The
line in (a) is the result of TMR plus LMR at 2 K. }\vspace*{-5mm}
\end{figure}
Magnetoresistivity ($MR=[\rho(H)-\rho(0)]/\rho(0)$) was measured
by sweeping the field at fixed temperature. Variation of the
in-plane and out-of-plane MR with H at fixed temperatures for
x=0.5 sample are shown in Fig. 2. Figure 2(a) shows the in-plane
MR at 2 and 20 K under the fields with different angles relative
to ab plane. A large, negative MR is observed at 20 K with H in ab
plane and along I, while the MR is very small with $H\perp ab$
plane. At 2 K, a large, positive MR is observed with $H\perp ab$
plane, while a large, negative MR is observed when H is applied
within ab plane and parallel to I. However, the MR shows a
complicated behavior with an angle of 45 degree between H and I,
the MR is negative, and monotonically increases with H up to 3.5
T, then decreases to zero at about 7 T and changes the sign with
further increasing H. It implied that the MR is contributed by a
positive and a negative component. In order to understand it, we
calculate the result of MR with $H\perp ab$ adding MR with
$H\parallel ab$, shown as the solid line in Fig. 2(a). The
calculated result is almost the same as the experimental one
except for the high H regime. It further suggests that the MR with
an angle of 45 degree between H and I is a competing result
between the MR's with $H\perp ab$ plane and $H\parallel ab$ plane,
respectively. Usually, the transverse MR contains the orbital
contributions, involving the contribution from the "bending" of
electron trajectory by the Lorentz force, which is always
positive. But here the positive MR is not only from the "bending"
of electron trajectory, the main contribution should be from spin
related effect because the transverse MR at 20 K is almost zero.
If the transverse MR would originate from the conventional orbital
motion of carriers, the MR should follow the Kohler's rule, i.e.
$\Delta\rho_H/\rho_0$ vs $(H/\rho_0)^2$ forms a single universal
relation, independent of temperature\cite{pippard}. However, the
MR at 2 K and 20 K violates the Kohler's rule obviously. In
addition, the longitudinal MR at 20 K is nearly the same as that
at 2 K. Because low temperature MR is usually larger than that at
high temperature except for a phase transition, it further
suggests that the MR is dependent on spin ordering in low
temperatures, spin reorientation at 20 K\cite{huang}.

The isothermal out-of-plane magnetoresistivity is shown in Fig.
2(b). At 45 K, the longitudinal MR shows a complicated behavior,
and experiences a maximum at about 6 T, while the transverse MR
monotonically increases with H. At 20K, the longitudinal MR is
positive and increases with increasing H, a negative transverse MR
is observed. This behavior is similar to that of in-plane at 2 K
shown in fig. 2(a). It is found that both the in-plane and
out-of-plane MR below 20 K is positive with $H\perp ab$ plane,
while negative with H applied in ab plane. Another interesting
feature is that the transverse MR at 45 K is much larger than that
at 20 K. It is considered that the larger negative MR could arise
from the suppression of charge ordering at about 50 K by
H\cite{balicas}. These results have indicated that the MR is
strongly dependent on spin ordering. $\mu SR$ measurements by
Uemura et al. gave an evidence that an antiferromagnetic order
sets in at the onset of a metal-insulator transition in the x=0.5
system. They gave a possible picture that one of two
interpenetrating Co spin networks acquires a long-range order
below 53 K, followed by the other network established long-range
order below 20 K\cite{Uemura}. At this moment, the spin structure
in the x=0.5 system is not clear, a conclusive picture requires
neutron scattering studies. Therefore, it is difficult to clearly
explain the observed MR behavior. The MR results seem to give an
evidence that the spin correlation is ferromagnetic in ab planes,
while antiferromagnetic between ab planes.

In order to understand the strong field direction dependence of
the MR observed in Fig. 2, the detailed in-plane MR measurements
are performed upon rotating H of 8 T, and H, I and c-axis always
are kept in the same plane. Evolution of the MR with the angle
between H and I is shown in Fig. 3(a). In-plane MR at 20 K is
negative over all angle range. At $\alpha=0$ ($\alpha$ is the
angle between H and I), the in-plane MR is maximum, while a MR is
almost zero at $\alpha=90^0$. A quite different behavior shows up
at 2 K, a striking anisotropy with a \emph{"d-wave-like"} symmetry
is observed in Fig. 3(b). $\Delta\rho_{ab}/\rho_{ab}$ changes from
negative at $\alpha=0^o$ to positive at $\alpha=90^o$, passing
through zero at $\alpha=45^o$. It should be noted that at 2 K the
positive term extending the "+" arm at $\alpha=0^{o}$ is larger
relative to the "-" arm at $\alpha=90^{0}$.
\begin{figure}[t]
\centering
\includegraphics[width=17cm]{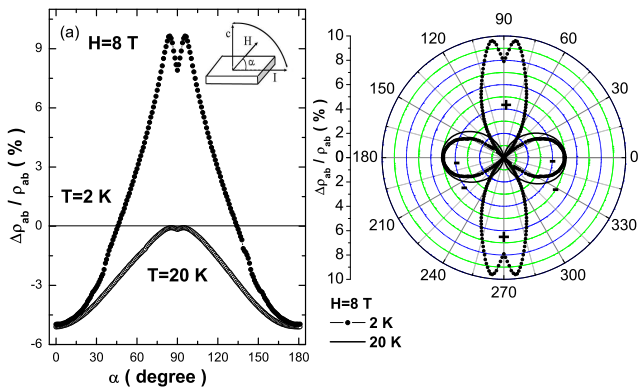}
\vspace*{-5mm} \caption{\label{fig:epsart}  The angular dependence
of in-plane magnetoresistivity for $Na_{0.5}CoO_2$ at H=8 T. The
angle between H and I is changed upon rotating H. H, I and c-axis
are kept in the same plane when the field is rotated.}
\vspace*{-5mm}
\end{figure}
Angular dependent MR at 20 K is also plotted in pole figure shown
in Fig. 3(b), the MR shows a two-fold symmetry. It should be
pointed out that the MR diagram in Fig. 3 is fairly symmetric at
20 and 2 K. The common behavior at the two temperatures is that
the negative arm appears at $\alpha=0^{o}$ and $180^o$, while the
difference is the absence of the positive arm at 20 K. It suggests
that the spin structure is changed with temperature decreasing
from 20 to 2 K, consistent with that observed by $\mu SR$
\cite{Uemura}, leading to the resistivity change at 20 K shown in
Fig.1(a) and (b). In x=0.5 system, the long-range
antiferromagnetic order occurred in two different Co spin networks
at 53 K and 20 K, respectively\cite{Uemura}, or the spin
reorientation took place at 20 K\cite{huang}. So far, we cannot
give an exact explanation about the anomalous "d-wave" shaped
angular dependent MR due to the unknown spin structure.  A similar
"d-wave-like" symmetric MR was observed below the Neel temperature
in antiferromagnetic $YBa_2Cu_3O_{6+x}$\cite{ando} and
$Nd_{2-x}Ce_xCuO_4$\cite{chen}. In $YBa_2Cu_3O_{6+x}$, the
magnetic field would give rise to a topological ordering of the
stripes, aligning them along the field direction and changing the
array of the current paths\cite{ando}. While in $Nd_2CuO_4$ the
spins prefer to be perpendicular to the field when the external
field is applied, so that the spin structure changes from
noncollinear to collinear\cite{skanthakumar}. In both of the
cases, the \emph{"d-wave"}-like anisotropy of in-plane MR is
understood as a consequence of the rotation of the stripe
direction with respect to the current direction\cite{ando} and of
the collinear spin direction relative to the current\cite{chen}.
We have no evidence for existence of a special spin structure in
the x=0.5 system, like the "spin stripes" ordering in
$YBa_2Cu_3O_{6+x}$ or "collinear spin structure" in $Nd_2CuO_4$,
but the anomalous angular dependence of MR implies existence of a
\emph{"spin ordering orientation"} turned by the field. Therefore,
neutron scattering measurement is acquired to resolve this
question.

In summary, transport properties and magnetoresistance were
studied in $Na_{x}CoO_2$ ($x=0.3$, $0.5$, and $0.7$). It is found
that $\rho_{ab}(T)$ shows similar temperature-dependent behavior
in $Na_{0.7}CoO_2$ and $Na_{0.3}CoO_2$, while $\rho_{c}(T)$ shows
different behavior. A dimensional crossover from two to three also
takes place with decreasing Na concentration. The angular
dependent in-plane magnetoresistivity for the x=0.5 system  is a
\emph{"d-wave-like"} symmetry at 2 K, while a \emph{"p-wave-like"}
symmetry at 20 K. Such anomalous angular dependence of MR is
related to a special \emph{"spin ordering orientation"} turned by
the applied field. Neutron scattering measurement under the
external field is needed to resolve this issue

This work is supported by the Nature Science Foundation of China
and by the Knowledge Innovation Project of Chinese Academy of
Sciences.

\end{document}